\documentclass[%
 reprint,
nofootinbib,
 amsmath,amssymb,
 aps,
 floatfix,
 tightenlines,
 prl,
]{revtex4-1}
\usepackage{dcolumn}
\usepackage{bm}
\usepackage{filecontents}
\usepackage{braket}
\usepackage{longtable}
\usepackage{color}

\usepackage{hyperref}

\begin{document}

\title{Lepton masses and mixing in a two-Higgs-doublet model}

\author{Piotr Chaber}%

\affiliation{Institute of Physics, University of Silesia, 75 Pu\l{}ku Piechoty 1, 41-500 Chorz\'{o}w, Poland}
\author{Bartosz Dziewit}
\affiliation{Institute of Physics, University of Silesia, 75 Pu\l{}ku Piechoty 1, 41-500 Chorz\'{o}w, Poland}
\author{Jacek Holeczek}
\affiliation{Institute of Physics, University of Silesia, 75 Pu\l{}ku Piechoty 1, 41-500 Chorz\'{o}w, Poland}
\author{Monika Richter}
\affiliation{Institute of Physics, University of Silesia, 75 Pu\l{}ku Piechoty 1, 41-500 Chorz\'{o}w, Poland}
\author{Marek Zra\l ek}
\affiliation{Institute of Physics, University of Silesia, 75 Pu\l{}ku Piechoty 1, 41-500 Chorz\'{o}w, Poland}
\author{Sebastian Zajac}
\affiliation{Faculty of Mathematics and Natural Studies, Cardinal Stefan Wyszynski University,  ul. Woycickiego 1/3, 01-938 Warsaw, Poland}

\date{\today}

\begin{abstract}
Within the framework of the two-Higgs Doublet Model (2HDM), we attempt to find some discrete, non-abelian flavour symmetry which could provide an explanation for the masses and mixing matrix elements of leptons. Unlike the Standard Model, currently there is no need for the flavour symmetry to be broken. With the \texttt{GAP} program we investigate all finite subgroups of the U3 group up to the order of 1025. Up to such an order there is no group for which it is possible to select free model parameters in order to match the masses of charged leptons, masses of neutrinos, and the Pontecorvo-Maki-Nakagawa-Sakata mixing matrix elements in a satisfactory manner.
\end{abstract}

\maketitle

\section{I. Introduction}

Despite the success of the Standard Model (SM) in providing a description of the current experimental data, there is a widespread belief that sooner or later an increase in available energy or accuracy of measurements will lead to the detection of a discrepancy between experimental results and theoretical predictions. The SM cannot be considered a complete theory because it does not provide answers to many pressing questions. One of the most important issues to be resolved concerns the masses of the fundamental constituents of matter, quarks and leptons. At the moment we still cannot theoretically predict their masses; we are only able to obtain their values from experimental data.

The discovery of the Higgs particle offers a partial solution to the problem, yet it does not resolve it completely. Particles acquire their mass by means of interaction with the Higgs field, and rather than examine the numerical values of the masses, we are currently more interested in the question of why particles interact so differently with the Higgs field, or in other words, why the Yukawa couplings cannot be theoretically predicted.

The solution to the problem of the elementary particles' masses is important in itself, because it would reduce the amount of unknown free parameters in the SM. Another important reason for conducting such investigations is that they create a great opportunity to understand the origin of the masses of physical bodies. The main part of the mass of each physical body comes from the interaction of the ingredients contained in it. But it is not the entire mass; the remaining part (though small) being the masses of individual fermions, which are still undetermined.

Several proposals to solve this problem, at least partially, can be found in the literature (see e.g. \cite{Ishimori:2010au,Blum:2007jz,Frampton:1994rk}). Although the problem concerns all matter constituents, here we will concentrate on an attempt to explain the masses and mixing angles of leptons. One of the most common approaches consists in the imposition of a flavour symmetry on the leptonic part of Yukawa Lagrangian (for review see e.g. \cite{King:2017guk, Altarelli:2010gt, King:2013eh}). This approach was particularly popular and successful before 2011 when it was discovered that the reactor-mixing angle $\theta_{13}$ is non-vanishing \cite{Harrison:2002er, Harrison:2003aw}. The models with an additional flavour symmetry are very popular, but they are by no means the only ones – some papers have been published in which the very existence of such a symmetry is denied  \cite{deGouvea:2012ac, Hall:1999sn}.

Attempts at solving the problem of lepton masses by a horizontal symmetry are dependent on the manner of the introduction of neutrino masses as well as on the Higgs sector for spontaneous symmetry breaking. In the simplest case of the conventional SM with one Higgs doublet in which neutrinos are massless, only three additional right-handed neutrinos are introduced. In such an extension of the SM, without introducing the Majorana term, neutrinos are Dirac particles \cite{Giunti:2007ry}. It is not necessary to introduce the right-handed neutrinos to obtain their masses. Instead, it is possible to use the existing left-handed neutrinos to form the Majorana masses \cite{Giunti:2007ry}. Both cases will be considered in this paper.

Within the framework of the Standard Model with one Higgs boson, a discrete symmetry for Yukawa couplings provides the relations for the three-dimensional mass matrices of charged leptons  $(M_l)$ and neutrinos $(M_\nu)$ \cite{Lam:2008sh, PhysRevLett.101.121602}:
\begin{eqnarray}
\label{eq1}
A^{i\dag}_L \left( M_l M_l^\dag \right) A_L^i &=& \left( M_l M_l^\dag \right),\\
\label{eq11}
A^{i\dag}_L \left( M_\nu M_\nu^\dag \right) A_L^i &=& \left( M_\nu M_\nu^\dag \right),
\end{eqnarray}
where $A^i_L=A_L(g_i), \;i=1,2,\ldots, N$ are three dimensional representation matrices for the left-handed lepton doublets for some N-order flavour symmetry group  $\mathcal{G}$.

In such a case, the I-st Schur’s lemma implies that $M_{l}M_{l}^{\dag}$ and $M_{\nu}M_{\nu}^{\dag}$ are proportional to the identity matrices, which clearly entails the trivial lepton mixing matrix (known in the literature as Pontecorvo-Maki-Nakagawa-Sakata (PMNS) matrix \cite{Pontecorvo:1967fh,ponton, Dziewit:2013aka}).

In order to avoid the relations given in (\ref{eq11}) and predict the non-trivial lepton masses and their mixing in this case, the family symmetry has to be broken. As a rule, the flavour symmetry $\mathcal{G}$ is spontaneously broken by scalar singlet Higgs fields called flavons (see e.g. \citep{King:2009ap, Altarelli:2010gt, King:2014nza}). However, it can also be broken by introducing a bigger number of normal Higgs multiplets (e.g. \cite{King:2014nza, Machado:2010uc}). The latter way is indeed more economical, since the spontaneous gauge symmetry breaking in this case gives rise to the particles’ masses, simultaneously leading to the break of a family symmetry. Additional flavon scalar fields in this framework are therefore redundant.

Such models were considered many times in the literature, but mostly in the frame of supersymmetric models where Higgs doublets were singlets of a flavour group \cite{Siyeon:2012bg, Nebot:2014nua}, or, in a more general approach, where only one selected flavour group was tested \cite{Kubo:2003iw, Ngu, Morisi:2009sc}.

In the present work we attempt to explore how the Two Higgs Doublet Model works in the context discussed above. As distinct from the previous works, we do not consider a few selected discrete groups, but instead we try to find a flavour symmetry in all groups up to 1025 order with one restriction, i.e. each of our groups must have at least one faithful, three dimensional irreducible representation.

In the next Section, we briefly introduce the flavour symmetry in the 2HDM model and show how the symmetry transformation between two Higgs doublets provides an opportunity to avoid the consequences of Schur's lemma. We also present all the formulas needed to conduct the computations in the case of Dirac and Majorana neutrinos. In Sec.~III the results of the final scan of the Yukawa matrices, the lepton masses and the PMNS mixing matrix elements are presented, and finally in Sec. IV we draw our conclusions.

\section{II. Two-Higgs-Doublet-Model with a flavour symmetry}

\subsection { A. Dirac neutrinos}

To begin, the leptonic part of the Yukawa Lagrangian with Dirac neutrinos will be considered. In contrast to the Standard Model, two Higgs doublets $\Phi_{i}$
contribute to the lepton masses (the so-called Two-Higgs-Doublet-Model of type III \cite{Hou:1991un}) as follows \cite{BRANCO20121}\footnote{Note that, in comparison to our notation, in the paper \cite{BRANCO20121}: $$\Phi_{i}=\left( \phi_{i}^{+}, \phi_{i}^{0}/\sqrt{2}\right)^T,\;i=1,2.$$}
\begin{eqnarray}\label{yukawa}
   \mathcal{L}_{Y}=&-&\sum\limits_{i=1,2}\sum\limits_{\alpha,\beta=e,\mu,\tau}\left( (h_{i}^{(l)})_{\alpha,\beta}\left[\overline{L}_{\alpha L}\tilde{\Phi}_{i}l_{\beta R}\right] \right.\nonumber \\
   &+& \left. (h^{(\nu)}_{i})_{\alpha\beta} \Big[\overline{L}_{\alpha L}\Phi_{i}\nu_{\beta R}\Big]\right)+h.c.
\end{eqnarray}
where:
$$L_{\alpha L}=\left(\begin{array}{c}\nu_{\alpha L}\\ \l_{\alpha L}\end{array}\right),\quad \Phi_{i}=\left(\begin{array}{c} \phi_{i}^{0}\\\phi_{i}^{-}\end{array}\right),\;i=1,2$$
are gauge doublets for the left-handed lepton and Higgs fields and the fields $l_{\beta R},\;\nu_{\beta R}$ stand for the right-handed lepton and neutrino fields, respectively.  The couplings $ h^{(l)}_{i}$ and $h^{(\nu)}_{i}$ create the 3-dimensional Yukawa matrices.

The spontaneous gauge symmetry breaking gives non-zero vacuum expectation values (VEVs) $v_{i}$ for the Higgs doublets:
\begin{equation}
\left<\Phi_{i}\right>=\frac{1}{\sqrt{2}}\left(\begin{array}{c}v_{i}\\0 \end{array}\right),
\end{equation}
and the mass matrices read as follows \cite{BRANCO20121}:
\begin{eqnarray}
M^{l}&=&-\frac{1}{\sqrt{2}}\left(v_{1}^{*}h^{(l)}_{1}+v_{2}^{*}h^{(l)}_{2}\right), \\
 M^{\nu}&=&\frac{1}{\sqrt{2}}\left(v_{1}h^{(\nu)}_{1}+v_{2}h^{(\nu)}_{2}\right).
\end{eqnarray}
In general the vacuum expectation values can be complex, $ v_{i}= \vert v_{i} \vert e^{i\varphi_{i}} $, but they are restricted by the Fermi coupling constant:
\begin{equation}
\label{Fermi}
\sqrt{\vert v_{1} \vert ^2 + \vert v_{2} \vert ^2} =\left( \sqrt{2} G_{F}\right)^{-1/2} \simeq  246 \ GeV.
\end{equation}

Family symmetry of our theory implies, that after the transformation of fields occurring in the 2HDM Lagrangian by the 3 dimensional ($A_{L},\;A^{R}_{l},\;A^{R}_{\nu}$) and 2 dimensional ($A_{\Phi}$) representations of  a flavour group $\mathcal{G}$:
\begin{eqnarray} \label{tran}
L_{\alpha L} \rightarrow L'_{\alpha L}&=& \left(A_{L}\right)_{\alpha, \chi} L_{\chi L},
\ l_{\beta R}\rightarrow l'_{\beta R}=\left(A^{R}_{l}\right)_{\beta, \delta} l_{\delta R} \nonumber \\
\nu_{\beta R}\rightarrow \nu'_{\beta R}&=&\left(A^{R}_{\nu}\right)_{\beta, \delta} \nu_{\delta R},\ \Phi_{i}\rightarrow \Phi'_{i}=\left(A_{\Phi}\right)_{ik}\Phi_{k},
\end{eqnarray}
the full 2HDM Lagrangian does not change:
\begin{equation}\label{inv}
\mathcal{L}\left(L_{\alpha L},l_{\beta R},\nu_{\gamma R},\Phi_{i}\right) =\mathcal{L}\left(L'_{\alpha L},l'_{\beta R},\nu'_{\gamma R},\Phi'_{i}\right).
\end{equation}

Given that all the transformation matrices in Eq.(\ref{tran}) are unitary, the only parts of the total 2HDM Lagrangian for which the aforementioned relations are not automatically fulfilled are the Yukawa Lagrangian and the Higgs potential. Imposition of symmetry on these terms of the model severely restricts their forms.

The invariance of the Yukawa Lagrangian is expressed as follows:
\begin{eqnarray}\label{inv2}
\mathcal{L}^{'}_{Y}\equiv &&-\sum_{i=1,2}\sum_{\alpha,\beta=e,\mu,\tau}\left( \left(h^{(l)}_{i}\right)_{\alpha,\beta}\left[\overline{L'}_{\alpha L}\tilde{\Phi'}_{i}l'_{\beta R}\right] \right. \nonumber \\
 && \left.+\left(h^{(\nu)}_{i}\right)_{\alpha\beta} \Big[\overline{L'}_{\alpha L}\Phi'_{i}\nu'_{\beta R}\Big]\right)+h.c.=\mathcal{L}_{Y}.
\end{eqnarray}

With regard to the Higgs potential, there appear to be two possibilities. The first one assumes that:
 \begin{equation}
 \label{SymmeYP}
 V\left(\Phi'_{1},\Phi'_{2}\right)=V\left(\Phi_{1},\Phi_{2}\right),
 \end{equation}
which implies that before and after the transformation for Higgs fields (Eq.(\ref{tran})) the coefficients in the potential remain exactly the same and the vacuum expectation values are equal, $v_i^\prime=v_i$.

There is also a second possibility, useful for phenomenological reasons, we allow for the modification of VEVs, which transform in the same way as the Higgs fields:
\begin{equation}
\label{transformation}
v'_{i}=\left(A_{\Phi}\right)_{ik}v_{k}.
\end{equation}

In this case, the form of the Higgs potential does not change, the terms in the potential do not vary, while only the potential coefficients undergo change.  This kind of invariance is known in the literature as the form-invariance (see \cite{BRANCO20121, PhysRevD.72.035004}).
After the unitary transformation (Eq.(\ref{transformation})), the condition given in Eq.(\ref{Fermi}) is unchanged, and hence also:
\begin{equation}
\label{Fermi1}
\sqrt{\vert v_{1}^{'} \vert ^2 + \vert v_{2}^{'} \vert ^2}  \simeq  246 \ GeV.
\end{equation}
The vacuum expectation values, whose sum of squares is constant, will need adjustments to meet the experimental requirements. Thus, from the point of view of flavour symmetry, the type of Higgs potential is irrelevant, so in our approach, the issue of what symmetry for Higgs Lagrangian is chosen becomes insignificant.

With reference  to the Eq.(\ref{inv}), in order to  find symmetric Yukawa matrices $h^{(l)}_{i},\;h^{(\nu)}_{i}$, $ i=1,2$, one can readily express the symmetry conditions  as the eigenequation for a direct product of unitary group representations to the eigenvalue 1 (see e.g.\cite{ludl1}):
\begin{widetext}
\begin{eqnarray}\label{relation}
\left((A_{\Phi})^{\dag} \otimes (A_{L})^{\dag} \otimes (A^{R}_{l})^{T}\right)_{k,\alpha,\delta;i,\beta,\gamma} (h^{l}_{i})_{\beta,\gamma}&=& (h^{l}_{k})_{\alpha,\delta}, \nonumber\\
\left((A_{\Phi})^{T} \otimes (A_{L})^{\dag} \otimes (A^{R}_{\nu})^{T}\right)_{k,\alpha,\delta;i,\beta,\gamma} (h^{\nu}_{i})_{\beta,\gamma}&=& (h^{\nu}_{k})_{\alpha,\delta},
\end{eqnarray}
\end{widetext}
for the charged leptons and neutrinos, respectively.

Both relations (Eq.(\ref{relation})) need to be  satisfied for any group's element $g\in \mathcal{G}$. It is however sufficient that they are fulfilled only for the group generators \cite{ludl1}, which considerably reduces the time of the computation.

In such a model, the invariance equations for the mass matrices are not trivial. For the symmetric Higgs potential (Eq.(\ref{SymmeYP})) :
\begin{equation}
A_{L}M^{l(\nu)}\left(A^{R}_{l (\nu)}\right)^{\dag}=\frac{1}{\sqrt{2}}\sum_{i,k=1}^{2} h^{l(\nu)}_{i} \left(A_{\Phi}\right)_{i,k}v_{k} \neq M^{l(\nu)}.
\end{equation}
then Eq.(\ref{eq1}-\ref{eq11}) are not satisfied and we avoid the consequences of the Schur's Lemma. The same can be shown for the form-invariant Higgs potential, where Eq.(\ref{transformation}) is satisfied. In this context we can obtain the non-trivial mass matrices without the introduction of additional flavon fields.

\subsection { B. Majorana neutrinos}

For Majorana neutrinos the Yukawa term has to be changed. In 2HDM, the simplest Yukawa Lagrangian can be taken as the non-renormalizable Weinberg term in the form:
\begin{eqnarray}\label{Maj}
\mathcal{L}_{Y}^{\nu}  \equiv &&-\frac{g}{M} \sum\limits_{i,k=1}^{2}\sum\limits_{\alpha,\beta=e,\mu,\tau} h^{(i,k)}_{\alpha,\beta}\left(\overline{L}_{\alpha L}{\Phi}_{i}\right)\left({\Phi}_{k} {L}_{\beta R}^{c}\right) \nonumber  \\
&&+  h.c,
\end{eqnarray}
where ${L}_{\beta R}^{c}=C\overline{L}_{\beta L}^{T}$
is the  charge conjugated lepton doublet fields. After the spontaneous symmetry breaking, the neutrino mass matrix is obtained:
\begin{equation}
\label{MassMatrix}
M^{\nu}_{\alpha,\beta}=\frac{g}{M}\sum_{i,k=1}^{2}v_{i}v_{k}h^{(i,k)}_{\alpha,\beta}.
\end{equation}
As in the preceding case, in compliance with the requirement of flavour symmetry for the Yukawa Lagrangian (Eq.(\ref{Maj})), the neutrino Yukawa matrices must satisfy the eigenvalue equation:
\begin{widetext}
\begin{eqnarray}
\label{InvEquation}
\left((A_{\Phi})^{T} \otimes \left(A_{\Phi}\right)^{T}\otimes\left(A_{L}\right)^{\dag} \otimes \left(A_{L}\right)^{\dag}\right)_{k,m,\chi,\eta;\ i,j,\alpha,\beta} \left(h^{(i,j)}_{\alpha,\beta}\right) = \left(h^{(k,m)}_{\chi,\eta}\right).
\end{eqnarray}
\end{widetext}
Such flavour symmetric Yukawa couplings restrict the neutrino mass matrix from Eq.(\ref{MassMatrix}).

\section{III. Results of the investigation for finite groups up to 1025 order }

\subsection{ A. The candidates for the flavour group \texorpdfstring{$\mathcal{G}_{F}$}{GF}}

The flavour group $\mathcal{G}_{F}$, which is imposed on the \textsl{2HDM} Lagrangian, cannot be arbitrary. Due to the fields' transformations defined in Eq.(\ref{tran}), the group must possess at least one 2-dimensional (for $A_{\Phi}$) and at least one 3-dimensional irreducible representation (for $A_{L}$, $A^{R}_{\nu}$ and $A^{R}_{l}$).  The sole application of irreducible representations is justified given the fact  that were any of the representations ($A_{\Phi}, A_{L}$ or  $A^{R}_{\nu}(A^{R}_{l})$) to be reducible, invariant Yukawa couplings would split up into independent sets for irreducible representations (see e.g.\cite{ludl1}).

In the selection of flavour symmetry groups, we have limited ourselves to finite-dimensional groups of the order of at most 1025, which are furthermore subgroups of the $U(3)$ group (at least one of the 3-dimensional irreducible or reducible representations must be faithful) \cite{ludl, PhysRevD.84.013011}. This additional condition is not necessary \cite{Grimus} but it significantly reduces the number of groups to be processed. Using the \texttt{GAP} (version 4.7.6) \cite{GAP4} system  for computational discrete algebra with the included \texttt{Small Groups Library} \cite{SmallGroup} and \texttt{REPSN} \cite{repsn} packages, we have found in total \texttt{10862} groups with at least one 2-dimensional and at least one 3-dimensional irreducible representation, but only \texttt{413} of these groups are subgroups of the $U(3)$ group. They split into two  disjoint sets. Each group has either at least one faithful 3-dimensional irreducible representation (there are \texttt{173} such groups), or at least one faithful 1+2 reducible representation (there are \texttt{240} such groups). Some groups are also subgroups of the $U(2)$ group. They have at least one faithful 2-dimensional irreducible representation (none of them have any faithful 3-dimensional irreducible representation). None of the groups have any faithful 1+1+1 reducible,  faithful 1+1 reducible or faithful 1-dimensional irreducible representation. All the groups which delivered any solutions belong to polycyclic groups that use the polycyclic presentation for element arithmetic (so called \textit{$PC$-groups}).

\begin{table*}
\caption{\label{tabelka} Groups of the order of at most \texttt{100} subject to consideration (all  the groups are listed in the ancillary file). Here:  ``$ [\ o\ ,\ i\ ] $'' the $i$-th group of the order $o$ in the \texttt{Small Groups Library} catalogue, ``\texttt{StructureDescription}'' a short string which provides some insight into the structure of the group under consideration, ``\texttt{2-D}'' the number of 2-dimensional irreducible representations, ``\texttt{3-D}'' the number of 3-dimensional irreducible representations, ``$U(2)$'' an indicator why the group is classified as a subgroup of the  $U(2)$ group (at least one 2-dimensional irreducible is faithful), ``$U(3)$'' an indicator why the group is classified as a subgroup of the  $U(3)$ group  (either at least one  3-dimensional irreducible  or one 1+2 reducible representation is faithful), ``\texttt{L}'' the number of different combinations of representations for charged leptons, ``\texttt{DN}'' the number of different combinations of representations for Dirac neutrinos, ``\texttt{MN}'' the number of different combinations of representations for Majorana neutrinos, ``\texttt{L+DN}'' the number of pairs of  different combinations of representations for charged leptons and Dirac neutrinos, ``\texttt{L+MN}'' the number of pairs of  different combinations of representations for charged leptons and Majorana neutrinos. Note that the ``\texttt{L}'' and the ``\texttt{DN}'' are always equal and that the ``\texttt{L+DN}'' is twice that number. All zero values are suppressed.}
\begin{ruledtabular}
\begin{tabular}{ccccccccccc}
 $ [\ o\ ,\ i\ ] $ & \texttt{StructureDescription} & 2-D  & 3-D& $U(2)$ & $U(3)$ & L & DN & MN & L+DN & L+MN\\ \hline
$[\  24 \ ,\ 3 \ ]$ & \texttt{ SL(2,3) } & 3 & 1 & 2 & 1+2 &   &   & 3 &   &   \\
$[\  24 \ ,\ 12 \ ]$ & \texttt{ S4 } & 1 & 2 &  & 3 & 4 & 4 & 2 & 8 & 4 \\
$[\  48 \ ,\ 28 \ ]$ & \texttt{ C2.S4=SL(2,3).C2 } & 3 & 2 & 2 & 1+2 & 4 & 4 & 6 & 8 & 4 \\
$[\  48 \ ,\ 29 \ ]$ & \texttt{ GL(2,3) } & 3 & 2 & 2 & 1+2 & 4 & 4 & 6 & 8 & 4 \\
$[\  48 \ ,\ 30 \ ]$ & \texttt{ A4:C4 } & 2 & 4 &  & 3 & 16 & 16 & 8 & 32 & 16 \\
$[\  48 \ ,\ 32 \ ]$ & \texttt{ C2xSL(2,3) } & 6 & 2 &  & 1+2 &   &   & 12 &   &   \\
$[\  48 \ ,\ 33 \ ]$ & \texttt{ SL(2,3):C2 } & 6 & 2 & 2 & 1+2 &   &   &   &   &   \\
$[\  48 \ ,\ 48 \ ]$ & \texttt{ C2xS4 } & 2 & 4 &  & 3 & 16 & 16 & 8 & 32 & 16 \\
$[\  54 \ ,\ 8 \ ]$ & \texttt{ ((C3xC3):C3):C2 } & 4 & 4 &  & 3 & 32 & 32 &   & 64 &   \\
$[\  72 \ ,\ 3 \ ]$ & \texttt{ Q8:C9 } & 9 & 3 & 2 & 1+2 &   &   & 9 &   &   \\
$[\  72 \ ,\ 25 \ ]$ & \texttt{ C3xSL(2,3) } & 9 & 3 & 2 & 1+2 &   &   & 9 &   &   \\
$[\  72 \ ,\ 42 \ ]$ & \texttt{ C3xS4 } & 3 & 6 &  & 3 & 36 & 36 & 6 & 72 & 12 \\
$[\  96 \ ,\ 64 \ ]$ & \texttt{ ((C4xC4):C3):C2 } & 1 & 6 &  & 3 & 12 & 12 & 2 & 24 & 4 \\
$[\  96 \ ,\ 65 \ ]$ & \texttt{ A4:C8 } & 4 & 8 &  & 3 & 64 & 64 & 16 & 128 & 32 \\
$[\  96 \ ,\ 66 \ ]$ & \texttt{ SL(2,3):C4 } & 6 & 4 & & 1+2 & 16 & 16 & 24 & 32 & 16 \\
$[\  96 \ ,\ 67 \ ]$ & \texttt{ SL(2,3):C4 } & 6 & 4 & 2 & 1+2 & 16 & 16 & 8 & 32 & 16 \\
$[\  96 \ ,\ 69 \ ]$ & \texttt{ C4xSL(2,3) } & 12 & 4 &  & 1+2 &   &   & 24 &   &   \\
$[\  96 \ ,\ 74 \ ]$ & \texttt{ ((C8xC2):C2):C3 } & 12 & 4 & 2 & 1+2 &   &   &   &   &   \\
$[\  96 \ ,\ 186 \ ]$ & \texttt{ C4xS4 } & 4 & 8 &  & 3 & 64 & 64 & 16 & 128 & 32 \\
$[\  96 \ ,\ 188 \ ]$ & \texttt{ C2x(C2.S4=SL(2,3).C2) } & 6 & 4 &  & 1+2 & 16 & 16 & 24 & 32 & 16 \\
$[\  96 \ ,\ 189 \ ]$ & \texttt{ C2xGL(2,3) } & 6 & 4 &  & 1+2 & 16 & 16 & 24 & 32 & 16 \\
$[\  96 \ ,\ 192 \ ]$ & \texttt{ (C2.S4=SL(2,3).C2):C2 } & 6 & 4 & 2 & 1+2 & 16 & 16 & 8 & 32 & 16 \\
$[\  96 \ ,\ 200 \ ]$ & \texttt{ C2x(SL(2,3):C2) } & 12 & 4 &  & 1+2 &   &   &   &   &   \\
\end{tabular}
\end{ruledtabular}
\end{table*}

All the detected groups of the order of at most \texttt{1025} are listed in the ancillary file, while Table \ref{tabelka} shows  groups of the order of at most \texttt{100} only which underwent investigation.

\subsection{B. Yukawa couplings matrices in the model with Dirac neutrinos}

In the case of all the groups subject to consideration, there exist 267 groups that gave in total 748672 different combinations of 2- and 3-dimensional irreducible representations that give 1-dimensional degeneration subspace for all generators, which is the solution to the equations in Eq.(\ref{relation}). This common vector gives Yukawa matrices for charged leptons ($h^{(l)}$) and for neutrinos ($h^{(\nu)}$), which are interrelated. All the possible solutions for Yukawa matrices for charged leptons and for Dirac neutrinos can be expressed through 7 base forms ($\omega=e^{2 \pi i/3}$):
\begin{eqnarray}
\label{OmegaR1}
h_{1}^{(1)} = \left(
\begin{array}{ccc}
   0 & 0 & 1 \\
  1 & 0 & 0 \\
  0 & 1 & 0 \\
\end{array}%
\right), \  \quad
h_{2}^{(1)} = \left(%
\begin{array}{ccc}
   0 & 1 & 0 \\
  0 & 0 & 1 \\
  1 & 0 & 0 \\
\end{array}%
\right),
\end{eqnarray}

\begin{eqnarray}
\label{OmegaR2}
h_{1}^{(2)} = \left(
\begin{array}{ccc}
   0 & 0 & 1 \\
  \omega^{2} & 0 & 0 \\
  0 & \omega & 0 \\
\end{array}%
\right), \quad
h_{2}^{(2)} = \left(%
\begin{array}{ccc}
   0 & 1 & 0 \\
  0 & 0 & \omega \\
  \omega^{2} & 0 & 0 \\
\end{array}%
\right),
\end{eqnarray}
\begin{eqnarray}
\label{OmegaR3}
h_{1}^{(3)} = \left(
\begin{array}{ccc}
   0 & 0 & 1 \\
  \omega & 0 & 0 \\
  0 & \omega^{2} & 0 \\
\end{array}%
\right), \quad
h_{2}^{(3)} = \left(%
\begin{array}{ccc}
   0 & 1 & 0 \\
  0 & 0 & \omega^{2} \\
  \omega & 0 & 0 \\
\end{array}%
\right),
\end{eqnarray}
the next three $h_{1}^{(i)}$ and $h_{2}^{(i)}$ for i = 4,5,6, are obtained from those given in Eq.(\ref{OmegaR1})-Eq.(\ref{OmegaR3}) by interchange, using the rule:
$$h_{1}^{(3+i)}=h_{2}^{(i)}, \ \   h_{2}^{(3+i)}=h_{1}^{(i)}$$ for i= 1,2,3, and finally\footnote{$U(3)$ subgroups which have no faithful 3-dimensional irreducible representation give only these diagonal solutions.}:

\begin{eqnarray}
\label{OmegaR7}
h_{1}^{(7)} = \left(
\begin{array}{ccc}
  1 & 0 & 0 \\
  0 & \omega^{2} & 0 \\
  0 & 0 & \omega \\
\end{array}%
\right), \quad
h_{2}^{(7)} = \left(%
\begin{array}{ccc}
   1 & 0 & 0 \\
  0 & \omega & 0 \\
  0 & 0 & \omega^{2} \\
\end{array}%
\right).
\end{eqnarray}

Symmetric Yukawa matrices for any considered group and for any irreducible representation within the group can be expressed now by seven (i=1,2,..,7) basic matrix forms as follows:\\
a) for Dirac neutrinos:
\begin{eqnarray}
\label{Neutrinos}
 \lbrace h_{1}^{(\nu)}, h_{2}^{(\nu)}\rbrace = \lbrace h_{1}^{(i)}, e^{i\varphi} h_{2}^{(i)}\rbrace,
 \end{eqnarray}
b) for charged leptons:
\begin{eqnarray}
\label{ChargedLeptons}
  \lbrace h_{1}^{(l)}, h_{2}^{(l)}\rbrace = \lbrace h_{2}^{(i)}, e^{-i(\delta_{l} +\varphi)} h_{1}^{(i)}\rbrace,
 \end{eqnarray}
where $\varphi$ is a phase distinctive for a group and for irreducible representations and $\delta_{l}=0,\pi.$\\
In order to find the lepton masses and mixing matrix we constructed the hermitian matrices as in Eq.(\ref{eq1}) $ M^{l}M^{l\dagger}$ and $ M^{\nu}M^{\nu\dagger}$. For all the possible Yukawa matrices we have obtained only three different forms $(x=l, \nu)$:
\begin{widetext}
\begin{eqnarray}\label{SquareofM}
M_{x}M_{x}^{\dagger} =  \vert c_{x}|^2
\left(
\begin{array}{ccc}
  1+\kappa^{2} & \kappa e^{-i(\eta_{x} + 2 k \pi/3)} & \kappa e^{i(\eta_{x} - 2 k \pi/3)} \\
  \kappa e^{i(\eta_{x} + 2 k \pi/3)} & 1+\kappa^{2} & \kappa e^{-i \eta_{x}} \\
  \kappa e^{-i(\eta_{x} - 2 k \pi/3)} & \kappa e^{i\eta_{x}}& 1+\kappa^{2} \\
\end{array} \right),
\end{eqnarray}
\end{widetext}
with $k=-1,0,+1$ and $\kappa = \vert v_{2}\vert/\vert v_{1}\vert$, the same for neutrinos and for charged leptons. The only difference lies in the phase $\eta_x$. For Dirac neutrinos, $$\eta_{\nu}= \varphi+\varphi_{2} -\varphi_{1},$$ and for charged leptons, $$\eta_{l} = \delta_{l}+\varphi +\varphi_{2}-\varphi_{1},$$
where $\varphi_i (i=1,2)$ are phases of the  VEVs $v_i$. After diagonalization of Eq.(\ref{SquareofM})
$$ U^{\dagger} \left(M_{x} M_{x}^{\dagger}\right) U= diag\left(m_{x1}^{2},m_{x2}^{2},m_{x3}^{2}\right),$$ we obtain:
\begin{eqnarray}
\label{mx1}
&m_{x1}^{2}&  =  |c_{x}|^2 \left(1+\kappa^{2} + 2  \kappa  cos\left(\eta_{x}\right)\right), \\
\label{mx2}
&m_{x2}^{2}&  = |c_{x}|^2 \left(1+\kappa^{2} + 2  \kappa  sin\left(\eta_{x}-\frac{\pi}{6}\right)\right),\\
\label{mx3}
&m_{x3}^{2}&  = |c_{x}|^2 \left(1+\kappa^{2} - 2  \kappa  sin\left(\eta_{x}+\frac{\pi}{6}\right)\right),
\end{eqnarray}
and the diagonalization matrix U:
\begin{equation}
\label{MatrixU}
U =    \frac{1}{\sqrt{3}}\left(
\begin{array}{ccc}
  e^{- \frac{2}{3} \pi i k} & \omega e^{- \frac{2}{3} \pi i k} & \omega^{2} e^{- \frac{2}{3} \pi i k} \\
  1 & \omega^{2} & \omega \\
 1 & 1 & 1 \\
\end{array} \right).
\end{equation}
This matrix does not depend on the phase $\eta_{x}$, so it is identical for charged leptons and for the neutrino. Therefore, it is not possible to reconstruct the correct mixing matrix. For groups and their irreducible representations for which $\delta_{l}= 0$, neutrinos and lepton masses are proportional and the mixing matrix is 3x3 identity matrix.
For groups and for representations where $\delta_{l}=\pi$, masses of charged leptons and neutrinos are not proportional. If the formulas in Eqs.(\ref{mx1}-\ref{mx3}) describe the masses of neutrinos, then for the charged leptons there is:
\begin{eqnarray}
\label{lx1}
&m_{l1}^{2} &=  |c_{l}|^2 \left(1+\kappa^{2} - 2  \kappa  cos\left(\eta_{\nu}\right)\right),\\
\label{lx2}
&m_{l2}^{2} &=  |c_{l}|^2 \left(1+\kappa^{2} - 2  \kappa  sin\left(\eta_{\nu}-\frac{\pi}{6}\right)\right),\\
\label{lx3}
&m_{l3}^{2} &=  |c_{l}|^2 \left(1+\kappa^{2} + 2  \kappa  sin\left(\eta_{\nu}+\frac{\pi}{6}\right)\right).
\end{eqnarray}
Similarly in this case we cannot reconstruct the PMNS mixing matrix for which the 3x3 anti-diagonal identity matrix is obtained.
 Regardless of whether the masses of charged leptons are described by formula Eqs.(\ref{mx1}-\ref{mx3}))  (groups with $\delta_{l} =0$) or by Eq.(\ref{lx1}-\ref{lx3}) (groups with $\delta_{l} = \pi$), three parameters $(c_{l}, \kappa,  \eta_{x}) $ cannot be selected in such a way as to obtain the physical masses of the electron, muon and tau (in any case, one obtains:  $0 \leqslant m_{e} / m_{\mu} \leqslant 1$ and $1 \leqslant m_{\tau} / m_{\mu} \leqslant 2$).

The SM extended by one additional doublet of Higgs particles (2HDM) does not posses a discrete family symmetry (in the groups under examination) that can explain the masses of charged leptons, masses of neutrinos having the nature of Dirac particles and the PMNS matrix.

\subsection{C. Yukawa couplings matrices in the model with Majorana neutrinos}

In the previous subsection, we did not receive masses of charged leptons consistent with the experiment. This was the case when the neutrinos were Dirac particles. In the case of Majorana neutrinos, we must once again look for a possible symmetry, since a symmetric solution for the Weinberg component may also deliver other groups of symmetry for charged leptons. We are currently looking for symmetries satisfying the Eg.(\ref{relation}) for the charged leptons  and  Eq.(\ref{InvEquation}) for neutrinos.
Out of all the groups under consideration, there exist 195 groups that gave in total 20888 solutions. All the found symmetries, as a solution of  Eq.(\ref{InvEquation}), give a two dimensional space which is  common to all generators of the groups in question. Each time we find two 36-dimensional vectors \textbf{p} and \textbf{r} and any linear combination of \textbf{p} and \textbf{r},  gives symmetric Yukawa matrices for Majorana neutrinos.
Thus these Yukawa matrices are given by:
\begin{equation}
\label{MajMass}
h^{(i,k)} = x p_{i,k} + y r_{i,k},
\end{equation}
where x and y are two free complex numbers. \\
A detailed analysis of the solution for Majorana neutrinos is as follows:
\begin{eqnarray}\label{MajMass2}
&h^{(1,1)}= x h_{2}^{(7)},\quad &h^{(1,2)}= y I_{3}, \nonumber \\
&h^{(2,1)}= y e^{i \delta} I_{3},\quad  &h^{(2,2)}= x e^{i(\delta + 2\varphi)} h_{1}^{(7)},
\end{eqnarray}
and for the  charged leptons Yukawa matrices:
\begin{eqnarray}
\label{LepMass}
\lbrace h_{1}^{(l)}, h_{2}^{(l)}\rbrace = \lbrace h_{2}^{(7)}, e^{-i( \delta_{l} +\varphi)} h_{1}^{(7)}\rbrace,
\end{eqnarray}
where $h_{1}^{(7)}$,  $h_{2}^{(7)}$ are given by Eq.(\ref{OmegaR7}) and $I_{3}$ is 3x3 identity matrix. As previously observed, the phases $\delta =(0,\pi), \delta_{l}$ and $\varphi$ depend on the group and its representations.The resulting neutrino mass matrix Eq.(\ref{MassMatrix}) has the form;
\begin{eqnarray}\label{MassMatrixR}
M^{\nu} =&&\frac{g}{2 M}\left(x \vert v_{1}\vert ^{2} e^{2 i \varphi_{1}}h_{2}^{(7)}+ y \vert v_{1} v_{2}\vert e^{i(\varphi_{1} + \varphi_{2})} I_{3} \left(1+e^{i\delta}\right) \right. \nonumber  \\
&&+ \left. x \vert v_{2}\vert ^{2} e^{i(\delta+2(\varphi_{2}+\varphi))}h_{1}^{(7)}\right),
\end{eqnarray}
which gives the squares of neutrino masses:
\begin{widetext}
\label{Mass21}
\begin{eqnarray}
m_{1}^{2}=&&\vert c_{\nu}\vert^{2} \left[ 1 +\kappa ^4  +2
   \beta ^2 \kappa ^2 + 2 \beta ^2 \kappa ^2 \cos (\delta )+4 \beta  \kappa ^3
   \cos \left(\frac{\delta }{2}\right)
   \cos \left(\frac{\delta }{2}+\tau +2
   \varphi \right) \right.\nonumber \\
   && \left. +4 \beta  \kappa  \cos 
   \left(\frac{\delta }{2}\right) \cos
   \left(\frac{\delta }{2}+\tau \right)+2
   \kappa ^2 \cos (\delta +2 \tau
   +2 \varphi ) \right],
\end{eqnarray}
\begin{eqnarray}
m_{2}^{2}=&&\vert c_{\nu}\vert^{2} \left[ 1  +\kappa
   ^4 +2
   \beta ^2 \kappa ^2+  2 \beta ^2 \kappa ^2 \cos (\delta )+4 \beta  \kappa ^3
   \cos \left(\frac{\delta }{2}\right)
   \sin \left(\frac{\delta }{2}+\tau +2
   \varphi -\frac{\pi }{6}\right) \right. \nonumber \\    
  &&\left. +4 \beta 
   \kappa    \cos \left(\frac{\delta
   }{2}\right) \sin \left(\frac{\delta
   }{2}+\tau -\frac{\pi }{6}\right)-2
   \kappa ^2 \sin \left(\delta +2 \tau +2
   \varphi +\frac{\pi }{6}\right) \right], 
\end{eqnarray}
\begin{eqnarray}
m_{3}^{2}=&&\vert c_{\nu}\vert^{2} \left[1+ \kappa
   ^4+2 \beta ^2 \kappa ^2+2 \beta ^2 \kappa ^2
   \cos (\delta )-4 \beta   \kappa
   ^3 \cos \left(\frac{\delta }{2}\right)
   \sin \left(\frac{\delta }{2}+\tau +2
   \varphi +\frac{\pi }{6}\right) \right. \nonumber \\
    && \left.-4 \beta
   \kappa  \cos \left(\frac{\delta
   }{2}\right) \sin \left(\frac{\delta
   }{2}+\tau +\frac{\pi }{6}\right)+2
   \kappa ^2 \sin \left(\delta +2 \tau +2
   \varphi -\frac{\pi }{6}\right) \right],
\end{eqnarray}
\end{widetext}
where $\beta=y/x$ , $\tau=\varphi_{2} - \varphi_{1}$ and $c_{\nu}= g x \vert v_{1}\vert ^{2} e^{2 i \varphi_{1}}/(2 M)$.

The masses of charged leptons are given by the same formulae as given above  Eqs.(\ref{lx1}-\ref{lx3}) and it is impossible to fit electron, muon and tau lepton masses. Our mass matrices for  neutrinos and charged leptons are diagonal, so we also do not have the ability to fit the mixing matrix.

We also see that the symmetry condition for Majorana neutrinos (Eq.\ref{InvEquation}) does not give  any new flavour symmetry group with new three dimensional representation $A_L$ which would not be present in the symmetry equations for charged leptons (Eq.\ref{relation}). The Yukawa matrices $h_1^{(l)}$ and $h_2^{(l)}$ in Eq.\ref{LepMass} for charged leptons are exactly the same as in the case of Dirac neutrinos (Eq.\ref{ChargedLeptons}). This is not a good conclusion. In the set of groups we are consider, regardless of the adopted neutrino sector, we will not find a symmetry that gives acceptable solutions for mass of charged leptons.

\section{IV. Conclusions}
We have explored the possibility of using some discrete  flavour symmetry to explain the masses and mixing matrix elements of leptons in the Standard Model with the Higgs particle sector extended by one additional Higgs doublet - 2HDM. In general, we have assumed that the total
Lagrangian model has a full flavour symmetry with one exception. We also admit that the Higgs potential in our model is only invariant in form, which is often used in the model  description of experimental data.  In such a model, we have avoided having to break the family symmetry and introduce flavon fields.\\
We have investigated discrete groups that are subgroups of the continuous $U(3)$ group up to the order of 1025. Models in which neutrinos have the nature of Dirac particles and models with Majorana neutrinos have been considered. Following a close analysis of these 413 groups and all their possible combinations of 2- and 3-dimensional irreducible representations, it is established  that none of them can reproduce the current experimental data.
Thus, in the 2HDM model with  the symmetrical or only form-invariant Higgs potential, in the class of the groups under consideration, there is no discrete family symmetry that would fully clarify the masses and parameters of the mixing matrix for leptons. In addition, we have observed that the set of symmetric Yukawa matrices for charged leptons is independent from the nature of the neutrinos. This can serve as a guidance for further search for family symmetry. In the models with two Higgs particles, regardless of the adopted neutrino sector and the set of groups that we consider, we do not find a symmetry that gives real masses of charged leptons, even approximately.

\begin{acknowledgments}
This  work  has  been  supported  by  the Polish~National~Science~Centre (NCN) under grant no.~UMO~-~2013/09/B/ST2/03382
\end{acknowledgments}
\bibliography{bibliography}
\end{document}